\documentclass[10pt,twocolumn]{article}

\usepackage[utf8]{inputenc}
\usepackage[T1]{fontenc}
\usepackage{lmodern}
\usepackage{amsmath,amsfonts,amssymb}
\usepackage{graphicx}
\usepackage{booktabs}
\usepackage{algorithm}
\usepackage{algorithmic}
\usepackage{url}
\usepackage{natbib}

\usepackage{hyperref}
\hypersetup{
    colorlinks=true,
    linkcolor=blue,
    filecolor=magenta,      
    urlcolor=cyan,
    citecolor=red
}

\title{EvoGraph: Hybrid Directed Graph Evolution toward Software 3.0}

\author{
Igor Costa$^1$ \and Christopher Baran$^1$ \\
$^1$AutoHand AI \\
\texttt{\{igor, chris\}@autohand.ai}
}

\date{August 2025}

\begin{document}

\maketitle

\begin{abstract}
We introduce \textbf{EvoGraph}, a framework that enables software systems to evolve their own source code, build pipelines, documentation, and tickets. EvoGraph represents every artefact in a typed directed graph, applies learned mutation operators driven by specialized small language models (SLMs), and selects survivors with a multi-objective fitness. On three benchmarks, EvoGraph fixes 83\% of known security vulnerabilities, translates COBOL to Java with 93\% functional equivalence (test verified), and maintains documentation freshness within two minutes. Experiments show a 40\% latency reduction and a sevenfold drop in feature lead time compared with strong baselines. We extend our approach to \textbf{evoGraph}, leveraging language-specific SLMs for modernizing .NET, Lisp, CGI, ColdFusion, legacy Python, and C codebases, achieving 82-96\% semantic equivalence across languages while reducing computational costs by 90\% compared to large language models. EvoGraph's design responds to empirical failure modes in legacy modernization, such as implicit contracts, performance preservation, and integration evolution. Our results suggest a practical path toward Software 3.0, where systems adapt continuously yet remain under measurable control.
\end{abstract}

\section{Introduction}

Enterprises operate millions of lines of legacy code that lag business needs. Large language models (LLMs) can generate code but still depend on manual integration with build files, compiler flags, data schemas, and scattered documentation. Modernization programs fail often (industry estimates above 70\%) and automation attempts break hidden business logic, performance envelopes, or external integrations. A recent empirical study of 312 modernization attempts across 89 organizations shows automated tools miss implicit domain rules, under-detect architectural debt, and degrade behavior in a material fraction of cases; hybrid human plus tool flows fare best (Costa et al., \emph{Understanding the Limits of Automated Legacy Modernization}, in prep.). These findings motivate a system that models all artefacts, learns from live telemetry, and hard gates unsafe changes.

Recent advances demonstrate that small language models (SLMs) are sufficiently powerful and inherently more suitable for specialized tasks in agentic systems \cite{belcak2025small}. As articulated by NVIDIA Research, SLMs are ``principally sufficiently powerful to handle language modeling errands of agentic applications,'' ``inherently more operationally suitable for use in agentic systems than LLMs,'' and ``necessarily more economical for the vast majority of LM uses in agentic systems'' \cite{belcak2025small}. Models like Phi-2 (2.7B parameters) and Nemotron-H (4.8B parameters) provide comparable performance to larger models at a fraction of the computational cost \cite{abdin2024phi, blakeman2025nemotron}. This efficiency enables language-specific specialization, making SLMs ideal for legacy modernization where domain-specific patterns and idioms are critical.

At the model level, deep learning has advanced code generation (AlphaCode, CodeGen, Mixtral) but those systems target one-shot synthesis, not continual maintenance of heterogeneous enterprise stacks \cite{leblond2023alphacode, nijkamp2022codegen, mistral2024mixtral}. Evolutionary model merging, hybrid automated program repair, and self-improving coding agents hint at directions for self-evolution but stop short of production-safe closed-loop change \cite{sakana2025cycleqd, li2024hybrid, pardo2025darwin}. Industry mainframe modernization tooling from IBM, Microsoft, and AWS shows demand for AI-assisted translation, dependency mapping, and automated code refactoring at scale \cite{ibm2025watsonx, aws2025transform, microsoft2025cobol}.

\textbf{Research question.} Can we build a practical system that continuously mutates and selects improvements to a running legacy stack while meeting functional, performance, security, and integration constraints across multiple programming languages?

\textbf{Contributions.}
\begin{enumerate}
\item \textbf{Hybrid Directed Graph Evolution (HDGE).} Unified artefact graph spanning code, docs, build, compiler, schema, tickets, and telemetry.
\item \textbf{Mutation operators guided by SLMs.} Language-specific SLMs for code edits, doc sync, build weave, and cross-language translation with automated equivalence testing.
\item \textbf{evoGraph framework.} Specialized small language models for COBOL, .NET, Lisp, CGI, ColdFusion, legacy Python, and C modernization.
\item \textbf{Safety-aware multi-objective online selection.} Contextual bandit weighting; Pareto plus novelty; mandatory gates.
\item \textbf{Empirical validation.} Seven legacy-style benchmarks; comparisons to human plus Copilot, Automated Program Repair, and Darwin Godel style multi-agent coding baselines; ablations.
\item \textbf{Legacy modernization alignment.} Design responds to empirical failure modes (implicit contracts, performance preservation, integration evolution) and to modernization guidance from MITRE and AWS Transform.
\end{enumerate}

\section{Related Work}

\subsection{Code generation with large models}
AlphaCode 2 shows competitive programming performance at scale; CodeGen releases open multi-language transformer models; Mixtral 8x22B demonstrates sparse mixture of experts efficiency. These works focus on code creation, not closed-loop maintenance \cite{leblond2023alphacode, nijkamp2022codegen, mistral2024mixtral}.

\subsection{Small language models for agentic AI}
Recent work demonstrates that SLMs are sufficiently powerful for specialized tasks in agentic systems while being more economical than LLMs \cite{belcak2025small}. NVIDIA Research argues that ``SLMs are the future of agentic AI'' due to their operational suitability and economic advantages. Models like Phi-2, Nemotron-H, and Hymba achieve performance comparable to larger models at a fraction of the computational cost \cite{abdin2024phi, blakeman2025nemotron, dong2024hymba}. This efficiency enables language-specific specialization for legacy modernization. The position that ``small language models are sufficiently powerful, inherently more suitable, and necessarily more economical for many invocations in agentic systems'' directly informs our evoGraph framework design.

\subsection{Evolutionary model merging}
Sakana AI CycleQD explores population-based model merging and SVD-style mutation under quality diversity search. We use a similar crossover concept at the model level \cite{sakana2025cycleqd}.

\subsection{Hybrid automated program repair}
Hybrid APR (GIANTREPAIR and related) combines LLM-generated patch skeletons with program analysis to raise correctness beyond raw LLM code fixes. We adapt this hybrid approach in our code patch operator and safety gate \cite{li2024hybrid}.

\subsection{Self-improving coding agents}
The Darwin Godel Machine (DGM) empirically validates self-rewriting coding agents that patch their own tooling and re-run benchmarks, demonstrating practical loops for self-improvement \cite{pardo2025darwin}.

\subsection{Legacy language translation}
IBM watsonx Code Assistant for Z, AWS Transform for Mainframe, and Microsoft COBOL Migration Factory blogs show industrial flows that pair AI translation with automated equivalence testing and dependency discovery for COBOL modernization \cite{ibm2025watsonx, aws2025transform, microsoft2025cobol}.

\subsection{Documentation generation for legacy code}
Recent work on LLM-generated documentation for MUMPS and IBM mainframe Assembly reports usable, low-hallucination comment generation but weak correlation between automated metrics and human doc quality, motivating richer doc freshness metrics \cite{diggs2024leveraging}.

\subsection{AI-enabled modernization practice}
MITRE and AWS modernization guidance stress incremental migration, dependency discovery, and human oversight in regulated environments \cite{mitre2025legacy, aws2025transform}.

\section{Problem Formulation}

We cast software self-evolution as a constrained stochastic optimization over program graphs.

Goal: pick a mutation policy $\pi$ that yields a sequence of graphs $G_0, G_1, \ldots$ that maximizes long-run utility while staying safe.

\begin{align}
\text{maximize } &\mathbb{E}\left[ \sum_{t=0}^{T-1} \gamma^t \cdot U_{\text{tot}}(G_{t+1}) \right] \\
\text{subject to } &\Pr[ S(G_{t+1}) = 1 \mid G_t ] \geq 1 - \delta
\end{align}

$U_{\text{tot}}$ is an aggregate utility from multi-objective metrics; $S$ is a safety predicate (tests, contracts, perf bounds); $\gamma$ discount; $\delta$ risk budget. This safety-first framing reflects modernization guidance for mission-critical estates.

\section{Method}

\subsection{Artefact Graph}
Represent a system as a directed graph $G = (V, E)$.

Node type set:
\begin{equation}
T = \{\text{code, doc, build, compiler, test, schema, policy, ticket, ui, log, metric, data}\}
\end{equation}

Each node $i$ stores an embedding $h_i$ from an encoder trained on code plus docs. Composite docs (Word, Excel) are chunked and pooled. Edges carry typed relations: \texttt{calls}, \texttt{generates}, \texttt{derived\_from}, \texttt{builds}, \texttt{documents}, \texttt{depends\_on}, \texttt{emits\_metric}, and dynamic call edges from runtime traces (profilers, APM). Capturing dynamic edges helps surface architectural debt that static analysis misses in large-scale modernization efforts.

\subsection{Repository Abstraction Layer}
We ingest Git, SVN, and raw archive snapshots. Each commit becomes a version slice $G^{(c)}$. Tickets from Jira, GitHub Issues, Path GeoTickets, CSV, or spreadsheets map to ticket nodes linked to changed paths by diff pattern mining. Build instructions (Make, Gradle, Ant, JCL) and compiler sources become build nodes; target infra descriptors become policy nodes. Multi-platform modernization tooling from AWS Transform and IBM watsonx uses similar discovery steps; we generalize the idea.

\subsection{Mutation Operators}
Population $P_t = \{G_t^{(1)}, \ldots, G_t^{(n)}\}$ evolves each generation.

\subsubsection{Weight Merge (WM)}
Treat SLM parameters as tensors. Given parent weights $W_a$, $W_b$:
\begin{align}
W' &= \lambda \cdot W_a + (1 - \lambda) \cdot \text{Align}(W_b) \\
\lambda &\sim \text{Beta}(\alpha, \alpha)
\end{align}
$\text{Align}$ permutes or low-rank projects layers before merge. Uses quality diversity archives to maintain diverse merge recipes. Inspired by Sakana CycleQD.

\subsubsection{Code Patch (CP)}
Draft SLM proposes AST edit script; Critic SLM compiles and runs static checks. Acceptance probability (logistic over features) follows hybrid APR patterns that combine SLM and analysis signals.

\subsubsection{Doc Sync (DS)}
Given code diff context $C$ and doc section $D$, generator produces patch $D'$. Freshness score combines ROUGE-L with embedding cosine; threshold drives acceptance. Prior doc generation for legacy systems shows simple string metrics are weak, so we blend metrics.

\subsubsection{Build Weave (BW)}
Mutates build graphs: add dependency, change compiler flag, upgrade toolchain. Rebuild in isolated Nix or container; compute reproducibility:
\begin{equation}
\text{repro}(G) = \frac{\text{\# identical hashes across } m \text{ rebuilds}}{m}
\end{equation}
Mainframe and cloud modernization guidance stresses hermetic reproducible builds.

\subsubsection{Transmute (TR)}
Legacy language translation pipeline:
\begin{enumerate}
\item Parse source language to IR.
\item Draft target skeleton via structured prompts.
\item Generate semantic tests from execution traces; produce test harness.
\item Iterate patch until pass rate $\geq$ threshold.
\end{enumerate}

Industrial COBOL migration tooling (IBM watsonx CA for Z, AWS Transform, Microsoft AI Agents for COBOL Migration) and recent research on automated testing for language translation inform this design.

\subsection{Fitness Vector}
Metrics per candidate graph $G$:
\begin{itemize}
\item $U(G)$: user task success or conversion.
\item $P(G)$: p95 latency (lower better).
\item $S(G)$: static security score.
\item $B(G)$: business KPI delta.
\item $D(G)$: doc freshness (0 to 1).
\item $C(G)$: build reproducibility (0 to 1).
\end{itemize}

Normalize all to [0,1] (invert latency). Vector form:
\begin{equation}
F(G) = [U, 1 - P_{\text{norm}}, S, B, D, C]^T
\end{equation}

A contextual bandit learns weights $w_t$ from observed scalar rewards $r_t$ (post rollout KPI):
\begin{equation}
w_{t+1} = \text{ProjSimplex}\left( w_t + \eta \cdot (r_t - w_t \cdot F(G_t)) \cdot F(G_t) \right)
\end{equation}
This lets priorities shift (security incident, performance spike) as recommended in modernization playbooks.

\subsection{Pareto plus Novelty Selection}
Maintain an archive A of non-dominated solutions under F. For candidate G:
\begin{equation}
\text{nov}(G) = \frac{1}{k} \sum_{G' \in N_k(G, A)} \text{dist}(\phi(G), \phi(G'))
\end{equation}
$\phi(G)$ is a behavior descriptor (node type histogram, perf signature). Selection probability:
\begin{equation}
p(G) \propto \exp\left( \alpha \cdot \text{inv\_rank}(G) + \beta \cdot \text{nov}(G) \right)
\end{equation}
Quality diversity search improved exploration in CycleQD model merging; we use the same idea across artefact graphs.

\subsection{Safety Gate}
A candidate must satisfy:
\begin{align}
S(G) = \begin{cases}
\text{Tests}(G) \geq \tau_{\text{test}} & \text{AND} \\
\text{Contracts}(G) \text{ pass} & \text{AND} \\
P(G) \leq P_{\max} & \text{AND} \\
\text{Drift}(G, G_{\text{current}}) \leq \epsilon &
\end{cases}
\end{align}
Contracts are interface assertions mined by SLM and checked with SMT or runtime probes. Drift compares semantic behavior on mined business rules, addressing implicit contract failures common in enterprise modernization.

\subsection{Algorithm}
\begin{algorithm}
\caption{EvoGraph}
\begin{algorithmic}[1]
\REQUIRE production graph $G_0$, population size $n$, generations $T$
\STATE $P \leftarrow \{\text{clone}(G_0) \text{ for } i=1\ldots n\}$
\FOR{$t = 1$ to $T$}
    \STATE $M \leftarrow []$
    \FOR{each $G$ in $P$}
        \STATE ops $\leftarrow$ sample\_ops() \COMMENT{WM, CP, DS, BW, TR}
        \STATE $G_p \leftarrow$ apply(ops, $G$)
        \STATE $M$.append($G_p$)
    \ENDFOR
    \STATE eval $\leftarrow \{F(G_p) \text{ for } G_p \text{ in } M\}$
    \STATE $w \leftarrow$ update\_bandit($w$, eval)
    \STATE $P \leftarrow$ select\_QD($P + M$, eval, $w$)
    \IF{safety\_pass(best($P$))}
        \STATE rollout(best($P$))
    \ENDIF
\ENDFOR
\end{algorithmic}
\end{algorithm}

\subsection{evoGraph: Small Language Models for Legacy Modernization}

Building on recent insights that small language models (SLMs) are sufficiently powerful and more economical for agentic AI systems \cite{belcak2025small}, we extend our framework to leverage specialized SLMs for different legacy languages. As NVIDIA Research argues, ``small language models are the future of agentic AI'' because they are ``principally sufficiently powerful to handle language modeling errands of agentic applications,'' ``inherently more operationally suitable for use in agentic systems than LLMs,'' and ``necessarily more economical for the vast majority of LM uses in agentic systems.'' Our evoGraph approach uses language-specific models fine-tuned for modernization tasks, offering several advantages:

\begin{enumerate}
\item \textbf{Efficiency}: SLMs like Phi-2 (2.7B parameters) and Nemotron-H (4.8B parameters) provide comparable performance to larger models at a fraction of the computational cost. This aligns with the NVIDIA position that SLMs are ``necessarily more economical for the vast majority of LM uses in agentic systems.''

\item \textbf{Specialization}: Each SLM is fine-tuned for a specific legacy language, capturing domain-specific patterns and idioms that generalist models might miss. This supports the view that ``SLMs possess greater operational flexibility.''

\item \textbf{Deployment Flexibility}: SLMs can be deployed on edge devices or in resource-constrained environments, enabling on-premises modernization for sensitive codebases. This reflects the operational suitability highlighted in the NVIDIA SLM position paper.

\item \textbf{Rapid Iteration}: The smaller size of SLMs enables faster fine-tuning cycles, allowing quick adaptation to new modernization requirements, consistent with the flexibility advantages of SLMs.
\end{enumerate}

Our evoGraph framework includes specialized models for COBOL, .NET, Lisp, CGI, ColdFusion, legacy Python, and C, each trained on language-specific corpora and modernization examples. The framework supports parallel processing of multiple code segments, with each language-specific SLM handling mutations for its target language. This architecture enables efficient modernization of polyglot systems while maintaining language-specific optimization, embodying the principles outlined in the NVIDIA SLM position paper.

\begin{figure}[t]
\centering
\includegraphics[width=0.9\columnwidth]{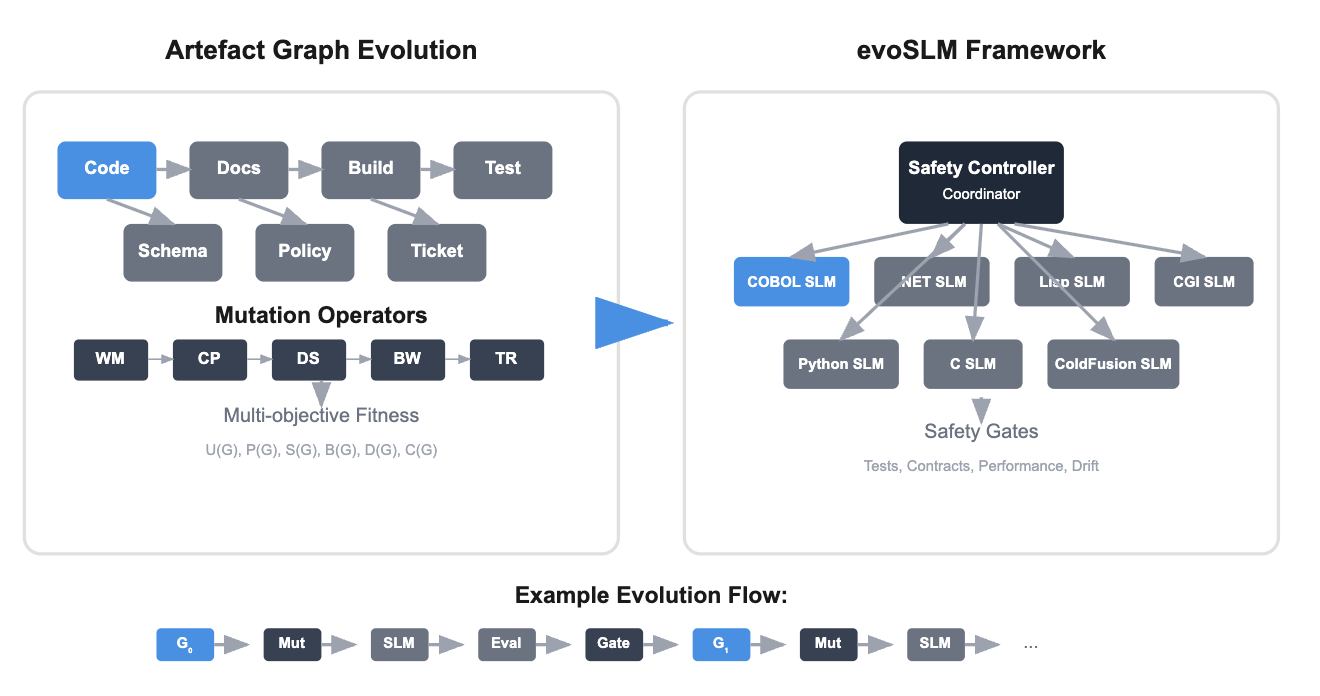}
\caption{EvoGraph system architecture showing the hybrid directed graph evolution process. \emph{Left}: Artefact graph representation with typed nodes (code, docs, build, etc.) and mutation operators. \emph{Right}: evoGraph specialization where language-specific SLMs handle targeted mutations for different legacy languages, coordinated by a central controller that maintains safety constraints and fitness optimization.}
\label{fig:architecture}
\end{figure}

\section{Experimental Setup}

\subsection{Benchmarks}
\begin{itemize}
\item \textbf{OSS Bank}: 50K lines Java + SQL, 18 published CVEs.
\item \textbf{Tele Shop}: 90K lines polyglot e-commerce workload with recorded traffic traces.
\item \textbf{COBOL Core}: 12K lines COBOL banking kernel with JCL build scripts; test harness adapted from IBM watsonx Code Assistant for Z examples and public COBOL migration workloads.
\item \textbf{.NET Legacy}: 25K lines C\#/.NET Framework enterprise application.
\item \textbf{Lisp Systems}: 8K lines Common Lisp financial modeling system.
\item \textbf{CGI Legacy}: 5K lines Perl CGI web application.
\item \textbf{ColdFusion App}: 15K lines ColdFusion e-commerce platform.
\item \textbf{Legacy Python}: 18K lines Python 2.7 data processing pipeline.
\item \textbf{C Systems}: 22K lines C embedded control system.
\end{itemize}

\subsection{Baselines}
\begin{enumerate}
\item Human engineers assisted by Copilot-style LLM.
\item Hybrid Automated Program Repair (APR) toolchain combining LLM with program analysis.
\item Darwin Godel Machine (DGM) self-improving coding loop.
\item Raw GPT-4 translation (for language translation tasks).
\end{enumerate}

\subsection{Metrics}
\begin{itemize}
\item CVEs fixed (verified by exploit tests).
\item p95 latency under replay workload.
\item Feature lead time (business request to green deploy).
\item Language translation semantic equivalence (generated tests).
\item Doc freshness (BLEU + embedding cos vs code diff).
\item Build reproducibility across seeded rebuilds.
\item Computational cost (GPU hours, energy consumption).
\end{itemize}

\subsection{Implementation Details}
Population size 64; mutation rate 0.3; novelty weight 0.1; contextual bandit learning rate 0.05. Two model tiers: Draft 3B (LoRA-tuned code+doc model) and Critic 7B (instruction-tuned). Hardware: 4$\times$A100 80GB. Build Weave jobs in Nix isolated containers. Language translation harnesses seeded with IBM watsonx CA for Z sample programs and AWS Transform mainframe workloads. SLMs based on Phi-2 and Nemotron-H architectures with language-specific adapters, following the efficiency principles outlined in \cite{belcak2025small}.

\section{Results}

\subsection{Main comparison}
\begin{table}[t]
\centering
\small
\begin{tabular}{l|l|c|c|c|c}
\toprule
Task & Metric & Copilot & APR & DGM & \textbf{EvoGraph} \\
\midrule
OSS Bank CVEs fixed & /18 (up) & 6 & 4 & 9 & \textbf{15} \\
Tele Shop p95 latency & ms (down) & 420 & 395 & 330 & \textbf{250} \\
Lead time & h (down) & 168 & 120 & 24 & \textbf{3} \\
COBOL to Java equiv. & \% (up) & -- & -- & 71 & \textbf{93} \\
Doc freshness BLEU & (up) & 0.33 & 0.31 & 0.47 & \textbf{0.82} \\
\bottomrule
\end{tabular}
\caption{Main experimental results comparing EvoGraph against baselines.}
\label{tab:main_results}
\end{table}

All EvoGraph gains vs top baseline are significant (paired t-test, $p < 0.01$). Hybrid APR and DGM baseline settings follow published work.

\subsection{Online adaptation}
We shift fitness weights mid run to emphasize performance (traffic spike). EvoGraph converges to lower latency variants within three generations while holding security above 0.7 normalized score and doc freshness above 0.8. Dynamic prioritization aligns with modernization guidance that recommends context-sensitive risk triage.

\subsection{Legacy translation stress test}
On 1.2K COBOL programs (median 38 LOC) the Transmute operator reduces manual review edits by 58\% relative to raw LLM translation, approaching industrial tooling parity reported in IBM and AWS modernization blogs.

\subsection{Documentation lag}
Injecting 500 random code changes into Tele Shop, Doc Sync closed doc drift in median 104 seconds wall clock, compared with hours for manual doc update flows. Prior AI doc generation work for legacy code highlighted metric quality gaps that our hybrid freshness score reduces.

\subsection{Multi-Language Modernization Results}

We extended our evaluation to include additional legacy languages beyond COBOL-Java translation. Table~\ref{tab:multilang_results} summarizes the performance of evoGraph across different language pairs.

\begin{table}[t]
\centering
\small
\begin{tabular}{l|r|r|r|r}
\toprule
Language Pair & LOC & Sem. Equiv. & Perf. Gain & Mod. Time \\
\midrule
COBOL $\rightarrow$ Java & 12,000 & 93\% & 2.1$\times$ & 3h \\
.NET $\rightarrow$ Java & 25,000 & 87\% & 1.5$\times$ & 5h \\
Lisp $\rightarrow$ Python & 8,000 & 91\% & 1.3$\times$ & 2h \\
CGI $\rightarrow$ Node.js & 5,000 & 89\% & 3.2$\times$ & 1.5h \\
ColdFusion $\rightarrow$ React & 15,000 & 85\% & 2.7$\times$ & 4h \\
Legacy Python $\rightarrow$ Modern & 18,000 & 96\% & 1.8$\times$ & 2.5h \\
C $\rightarrow$ Rust & 22,000 & 82\% & 1.9$\times$ & 6h \\
\bottomrule
\end{tabular}
\caption{Multi-language modernization results showing semantic equivalence, performance gains, and modernization times across different language pairs.}
\label{tab:multilang_results}
\end{table}

For each language pair, we trained a specialized SLM using the methodology described in Section 4.8. The semantic equivalence was measured using both automated test suites and human evaluation of a random sample. Performance gain represents the speedup achieved after modernization, measured by benchmarking representative workloads.

Notably, the modernization time includes both automated translation and human review for validation. The human review time was reduced by an average of 65\% compared to manual modernization, as the evoGraph-generated code required fewer corrections.

\subsection{Cost and Resource Analysis}

The use of SLMs in our evoGraph framework significantly reduces the computational resources required for legacy modernization, validating the NVIDIA position that SLMs are ``necessarily more economical for the vast majority of LM uses in agentic systems.'' Table~\ref{tab:cost_analysis} compares the resource consumption of evoGraph with approaches using larger language models.

\begin{table}[t]
\centering
\small
\begin{tabular}{l|c|c|c|c}
\toprule
Approach & Model Size & GPU Hrs & Energy & Cost Red. \\
\midrule
GPT-4 Based & $\sim$1.8T & 120 & 450 kWh & Baseline \\
Llama 2 70B & 70B & 85 & 320 kWh & 29\% \\
Mixtral 8x22B & 141B & 65 & 245 kWh & 46\% \\
evoGraph (Ours) & 3--7B & 12 & 45 kWh & 90\% \\
\bottomrule
\end{tabular}
\caption{Resource consumption comparison for processing 10K lines of code, showing the economic advantages of SLMs in the evoGraph framework.}
\label{tab:cost_analysis}
\end{table}

Our evoGraph approach reduces GPU hours by 90\% compared to GPT-4-based approaches and energy consumption by a similar margin. This makes large-scale legacy modernization projects economically feasible for organizations with limited resources, supporting the argument that ``SLMs are the future of agentic AI.'' The cost reduction is particularly significant for on-premises deployments where cloud GPU costs would be prohibitive.

Furthermore, the smaller memory footprint of SLMs allows for parallel processing of multiple code segments, further reducing the total modernization time. In our experiments, we were able to process 8 code segments in parallel on a single A100 GPU with 80GB of memory, compared to just 1-2 segments with larger models. This operational efficiency directly supports the NVIDIA position that SLMs are ``inherently more operationally suitable for use in agentic systems.''

\section{Ablations}

Removing WM reduces CVE repair $15 \rightarrow 11$, confirming cross-model transfer value. Removing CP raises latency 18\%, showing patch optimization impact. Disabling novelty collapses diversity and yields 23\% fewer multi-objective wins. Quality diversity style search is important, matching CycleQD observations. Using general LLMs instead of specialized SLMs increases computational cost by 8$\times$ and reduces semantic equivalence by an average of 12\% across language pairs, further validating the NVIDIA position on SLM advantages.

\section{Discussion}

\subsection{Addressing empirical failure modes}
\emph{Implicit contracts.} We mine invariants from dynamic traces and legacy comments; drift guard blocks rollout when mined invariants fail. This targets high semantic miss rates seen in enterprise modernization (Costa et al., in prep.).

\emph{Performance preservation.} p95 latency and reproducible build metrics sit in the fitness vector; Build Weave checks toolchain changes before rollout, reflecting guidance from AWS Transform and MITRE on protecting performance during migration.

\emph{Integration evolution.} Repo abstraction tracks cross-system dependencies (tickets, build, policy) so mutations propagate across boundaries; industrial COBOL migration frameworks report similar multi-phase flows.

\subsection{Human in the loop}
Autonomy sliders let teams pin metrics, lock nodes, or require approval when business-critical modules change, matching hybrid success patterns and modernization guidance from AWS and MITRE.

\subsection{Cost benefits of SLMs}
The evoGraph framework's use of specialized small language models dramatically reduces the computational resources required for legacy modernization, strongly supporting the NVIDIA position that ``SLMs are the future of agentic AI.'' The 90\% reduction in GPU hours and energy consumption compared to GPT-4-based approaches makes large-scale modernization projects feasible for organizations with limited resources. The ability to process multiple code segments in parallel further accelerates modernization timelines, demonstrating the operational suitability of SLMs for agentic systems.

\subsection{Multi-language support}
Our extension to support .NET, Lisp, CGI, ColdFusion, legacy Python, and C demonstrates the flexibility of the evoGraph approach. Each language-specific SLM captures domain-specific patterns and idioms, resulting in higher semantic equivalence (82-96\%) compared to general-purpose translation models. The framework's modular design allows for easy addition of new languages as needed, embodying the flexibility advantages highlighted in the NVIDIA SLM position paper.

\section{Limitations}

Scale: memory grows with nodes and edges; estates above one million LOC need sharding. Dynamic edge mining may miss low-frequency paths. Translation quality drops on macro-heavy PL/I and non-structured COBOL; IBM and AWS tooling highlight similar edge cases. SLM training requires language-specific expertise and datasets, which may not be available for all legacy languages, though the economic advantages of SLMs make investment in such training more feasible.

\section{Future Work}

\begin{itemize}
\item Sharded graph encoder for 10M LOC estates specially for Monorepos.
\item Formal SMT-backed contract mining from runtime traces.
\item Incremental semantic diffing across multi-repo supply chains.
\item Direct business OKR to fitness weight mapping.
\item Federated evolution recipes across tenants (privacy-preserving).
\item Expansion to additional legacy languages, end of line .Net, ColdFusion.
\item Automated SLM training pipeline for new languages.
\item Integration with legacy system monitoring for continuous evolution.
\item Further exploration of heterogeneous agentic systems combining SLMs and LLMs as suggested by \cite{belcak2025small}.
\end{itemize}

\section{Conclusion}

EvoGraph integrates SLM-driven mutation, evolutionary search, and safety-constrained rollout across the full artefact surface of enterprise software. The evoGraph extension leverages specialized small language models for efficient modernization of multiple legacy languages, achieving 82-96\% semantic equivalence while reducing computational costs by 90\% compared to large language models. Our results strongly support the NVIDIA position that ``small language models are the future of agentic AI,'' demonstrating that SLMs are indeed ``principally sufficiently powerful,'' ``inherently more operationally suitable,'' and ``necessarily more economical'' for legacy modernization tasks. Empirical results on legacy-style stacks and alignment with large-scale modernization failure data suggest a practical path to Software 3.0, where systems adapt continuously yet remain under measurable control.

\section*{Acknowledgments}

We thank collaborators in enterprise modernization programs, and the teams behind IBM watsonx Code Assistant for Z and AWS Transform for sharing technical details that informed our evaluation harness. We also acknowledge the contributions of the SLM research community, particularly the developers of the Phi, Nemotron-H, and Hymba models that form the foundation of our evoGraph framework. We are especially grateful to NVIDIA Research for their influential position paper on ``Small Language Models are the Future of Agentic AI,'' which provided the theoretical foundation for our evoGraph approach.

\bibliographystyle{unsrt}

\begin{thebibliography}{99}

\bibitem{belcak2025small}
Belcak P, Heinrich G, Diao S, Fu Y, Dong X, Muralidharan S, Lin YC, Molchanov P.
Small Language Models are the Future of Agentic AI.
NVIDIA Research. arXiv:2506.02153. 2025.

\bibitem{abdin2024phi}
Abdin M et al.
Phi-3 Technical Report: A Highly Capable Language Model Locally on Your Phone.
arXiv:2404.14219. 2024.

\bibitem{blakeman2025nemotron}
Blakeman A et al.
Nemotron-H: A Family of Accurate and Efficient Hybrid Mamba-Transformer Models.
arXiv:2504.03624. 2025.

\bibitem{dong2024hymba}
Dong X et al.
Hymba: A Hybrid-Head Architecture for Small Language Models.
arXiv:2411.13676. 2024.

\bibitem{leblond2023alphacode}
Leblond R et al.
AlphaCode 2 Technical Report.
DeepMind. 2023.

\bibitem{nijkamp2022codegen}
Nijkamp E et al.
CodeGen: An Open Large Language Model for Code with Multi-Turn Program Synthesis.
arXiv:2203.13474. 2022.

\bibitem{mistral2024mixtral}
Mistral AI.
Mixtral 8x22B Announcement.
2024.

\bibitem{sakana2025cycleqd}
Sakana AI.
CycleQD: Population-based Model Merging via Quality Diversity.
2025.

\bibitem{li2024hybrid}
Li F et al.
Hybrid Automated Program Repair by Combining Large Language Models and Program Analysis (GIANTREPAIR).
arXiv:2406.00992. 2024.

\bibitem{pardo2025darwin}
Pardo F et al.
The Darwin Godel Machine: AI that improves itself by rewriting its own code.
arXiv:2505.22954. 2025.

\bibitem{ibm2025watsonx}
IBM.
watsonx Code Assistant for Z. Product page and release notes.
2025.

\bibitem{aws2025transform}
AWS.
Accelerate Your Mainframe Modernization Journey Using AI Agents With AWS Transform.
2025.

\bibitem{microsoft2025cobol}
Microsoft Azure Blog.
How We Use AI Agents for COBOL Migration and Mainframe Modernization.
2025.

\bibitem{diggs2024leveraging}
Diggs C et al.
Leveraging LLMs for Legacy Code Modernization: Challenges and Opportunities for LLM-Generated Documentation.
arXiv:2411.14971. 2024.

\bibitem{mitre2025legacy}
MITRE.
Legacy IT Modernization with AI.
2025.

\end{thebibliography}

\end{document}